\documentstyle[twoside,fleqn,esp_dam,epsfig,amssymb]{article}

\newcommand{\AmS}{{\protect\the\textfont2
  A\kern-.1667em\lower.5ex\hbox{M}\kern-.125emS}}

\title{Decays of Heavy Mesons\thanks
{Talk given by D. Becirevic at ``Lattice99'', Pisa, Italy. He acknowledges the INFN for financial support.}}

\author{A.~Abada\address{Universit\'e de Paris Sud, L.P.T.~(B\^at.~210), 
91405 Orsay-Cedex, France}, D.~Becirevic\address{Dip. di Fisica, Univ. ``La Sapienza" and INFN, P.le A. Moro, I-00185 Rome, Italy}, Ph.~Boucaud$^{\rm a}$, J.P.~Leroy$^{\rm a}$, V.~Lubicz\address{Dip. di Fisica, Univ. di Roma Tre and INFN, Via della Vasca Navale 84, I-00146 Rome, Italy}, G.~Martinelli$^{\rm b}$, F.~Mescia$^{\rm b}$}       

\begin{document}
\newcommand{\sze}{\small}
\newcommand{\bea}{\begin{eqnarray}}
\newcommand{\eea}{\end{eqnarray}}
 \newcommand{\beq}{\begin{equation}}
\newcommand{\eeq}{\end{equation}}

\begin{abstract}
We present preliminary results for heavy to light transitions of pseudoscalar mesons, induced by the vector and tensor operators. This lattice study is performed in quenched approximation, by using the nonperturbatively improved Wilson action and operators. We also update the values of the heavy-light meson decay constants. 
\end{abstract}
 
\maketitle

The full ${\cal O} (a)$-improvement of the Wilson action and bilinear quark operators~\cite{alpha}, was tested in spectroscopy and proved to be efficient, namely the hadronic masses and decay constants indeed converge faster to their continuum limit values. Improved operators, including the ${\cal O} (a)$-free renormalization constants (RC's), have been used to determine the values of the light quark masses. The next important step is to use non-perturbative (NP) improvement for hadron$\to$hadron decays. Although, the light$\to$light meson decays represent natural candidates for such a study, it is phenomenologically more appealing to apply the improvement to the heavy$\to$light case. It is well known that these decays are subject to larger discretization errors  because the present values of the hard cutoff are $\sim 3$~GeV. To make predictions relevant for $B$-physics, one either relies on extrapolations in heavy masses~\cite{ukqcd}, or attempt to use effective theories~\cite{onogi_et_ryan}. On the contrary, $D$-meson decays can be studied directly by using non-perturbatively improved Wilson fermions. The effect of the removal of ${\cal O}(a)$-artifacts (in particular those $\sim a m$) is expected to be efficient in this case. Moreover, these (directly obtained) results can be used to test the predictions obtained using effective theories.

In this talk, we present our preliminary results concerning heavy to light transitions of pseudoscalar mesons. This includes:\\
\underline{\small{\sl \ Leptonic decays}} ($H\to \ell \nu_\ell$)
\vspace*{-1.mm}
\bea
\label{def0}
\langle 0  \vert \widehat A_0 \vert H(p_H)\rangle = i f_{H}  m_H\;,
\eea
\vspace*{-2.6mm}
\underline{\small{\sl Semileptonic decays}} ($H\to P\ell \nu_\ell$)
\vspace*{-1.mm}
\bea
\label{def1}
\langle P (p) \vert \widehat  V_\mu \vert H(p_H)\rangle = {m_H^2 - m_P^2 \over q^2} q_\mu
 F_0^{\scriptsize H\to P}(q^2) + \nonumber \\
 \left( p_H + p - q {m_H^2 -
m_P^2\over q^2} \right)_\mu  F_+^{ H\to P}(q^2) \ ,
\eea
\vspace*{-2.mm}
\underline{\small{\sl Penguin-induced decays}} ($H\to P\ell^+ \ell^-$)
\vspace*{-.7mm}
\bea
\label{def2}
&&\hspace*{-6.7mm}\langle P (p) \vert  \widehat T_{\mu}(\mu) \vert H(p_H)\rangle = \hfill \nonumber \\
&&\hspace*{-6.7mm}i
\left( q^2 (p_H + p) - (m_H^2
- m_P^2) q \right)_\mu \ {F_T(q^2)^{H\to P}\over m_H + m_P} \;, 
\eea
where $q=p_H-p$ is the momentum transferred from the heavy $H$-meson to the light $P$-pseudoscalar meson. Hat symbols indicate that the corresponding operator has been improved and renormalized. The scale $\mu$ is set to its standard(ized) value of $2$~GeV. As usual, the main goal is to compute the decay constant $f_{H}$, and the form factors $F_{0,+,T}(q^2)$ (the $\mu$-dependence in $F_{T}$ is implicit). In particular, in order to study the $q^2$-behavior of the form factors, one needs to inject various three-momenta to the interacting hadrons.

{\bf 1.~Lattice details.} To study the above decays, we produced a sample of $200$ (quenched) gauge field configurations on a $24^3\times 64$ lattice, at $\beta=6.2$. The NP-improved Wilson action has been used in calculating the propagators: we work with four light ($\ell$) and four heavy ($h$) quark species~\cite{heavy}. They are combined in calculation of relevant two- and three-point correlation functions. For the latter, the sources of light and heavy mesons are fixed at $t_P=0$ and $t_H=28$. The spectator quark is degenerate with the light one coming out from 
the main (transition) vertex. The axial ($A_\mu = \bar q \gamma_{\mu} \gamma_{5} Q$), vector ($V_\mu = \bar q \gamma_{\mu} Q$) and the penguin  
($T_\mu = i \bar q \sigma_{\mu \nu} q^\nu Q$) operators are improved by adding a divergence or curl of operators having the same discrete symmetry properties, with coefficients ($c_J$) tuned as to exactly cancel the ${\cal O}(a)$-artifacts. Similarly, the RC's are improved as $Z_J(g_0^2, \bar m)= Z_J^{(0)} (1 + b_J \bar m)$, where $\bar m = (m_h + m_\ell )/2$. In this study we use the nonperturbative quantities whenever available ($Z_A^{(0)}$, $Z_V^{(0)}$, $Z_T^{(0)}(\mu )$, $b_V$), otherwise the ones from the boosted perturbation theory ($c_{A}$, $c_{V}$, $c_{T}$, $b_T$). Higher order mass corrections can be handled at tree level, by multiplying the renormalization constants by the (modified) KLM factors~\cite{heavy}.\\
The lattice spacing is obtained by following the same procedure described in Ref.~\cite{light}. From the doubled statistics, and using the $K^*$-meson mass, we obtain $a^{-1}=2.72(11)$~GeV.

{\bf 2.~Decay constants.} We first update the values of Ref.~\cite{heavy}, for the heavy-light pseudoscalar meson decay constants:
\vspace*{-1.6mm} 
\bea
&&\hspace*{-6.7mm}f_{D}=216(11)^{+5}_{-4}\ {\rm MeV}\ ;
\;\; f_{D_s}=239(10)^{+15}_{-0}\ {\rm MeV}\ ;\cr
\vspace*{-.7mm} 
&& \cr
&&\hspace*{-6.7mm}f_{B}=173(13)^{+34}_{-2}\ {\rm  MeV}\ ;
\;\; f_{B_s}=196(11)^{+42}_{-0}\ {\rm MeV}\ ;\cr
\vspace*{-.7mm} 
&& \cr
&&\hspace*{-6.7mm} f_{D_s}/f_{D} = 1.11(1)^{+.01}_{-0}\ ;\;\;  f_{B_s}/f_{B} = 1.14(2)^{+.01}_{-.01}\ .
\nonumber
\eea 
\vspace*{-1.mm} 
\noindent
First errors are statistical, while the second are (conservative) systematic and include: the scale fixing uncertainty ({\it i.e.} the dispersion of the ratios with $m_{K^*}$, $f_K$, $f_\pi$), the RC's uncertainty (estimated by the difference with/without KLM factor), and difference between the quadratic and linear extrapolations in $1/M_H$ (which is the major source of error in $f_{B_{(s)}}$).
We also update the ratio:
\beq
f_{B}/f_{D_s} = 0.72(2)^{+.13}_{-0}\, ,
\eeq
\vspace*{-.7mm} 
which, when combined with $f_{D_s}^{(exp)}$, amounts to
\bea 
f_B = 183 \pm 22({\rm exp.})^{+30}_{-5}({\rm theo.})\ {\rm MeV}\ .
\eea

\begin{figure}[h!]
\vspace*{-1.1cm}
\hspace*{-.25cm}
\epsfig{file=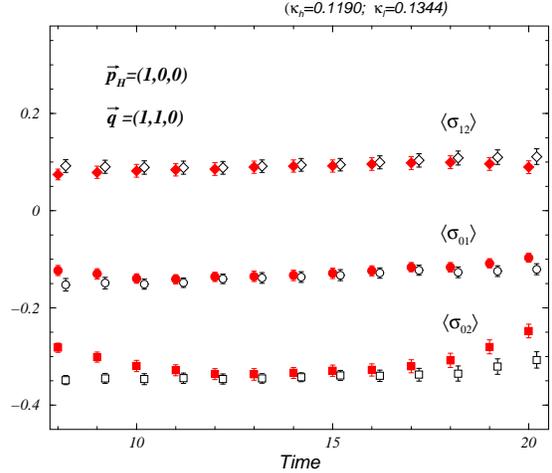,width=7.2cm} 
\vspace*{-1.15cm}
\caption{\footnotesize Extracted matrix element $\langle P\vert i \bar q  \sigma_{\mu \nu}Q \vert H \rangle$, before saturating by $q^\nu$. Filled (empty) symbols correspond to the matrix elements obtained from the numerical (analytic) ratio.}  
\label{fig1}  
\vspace*{-.8cm}
\end{figure}

{\bf 3.~Form factors.} We take advantage of the fact that we use the local source-operators. The $2^{pts}$-functions can then be fitted to the form
\bea
\label{disp}
{\cal C}^{(2)}(\vec p;t) = \left. {\cal Z}_P\right|_{\vec p = 0}\ { e^{- E_P t}  
\over 2 \ {\rm
sinh} E_P}\ ,
\eea
which allows us to reach the higher upper bound $\vert \vec p\vert < 3 $ (in units of elementary momentum on the lattice $2 \pi/La$), for which the latticized free boson dispersion relation is satisfied~\cite{light}.
We consider the heavy meson momenta $\vec p_H\in \{ (0,0,0); (1,0,0); (1,1,0)\}$ (and the cyclic permutations). That implies 32 nonequivalent $\vec q$, compatible with $\vert \vec p\vert <  3$. After inspection, it turns out that the signal for the form factors is `visible' for $15$ different momenta~\cite{prepa}. To extract the matrix element we examine the ratio: 
\bea
R_\mu(t) = {{\cal C}^{(3)}(\vec p, \vec q; t) \over \  {\cal C}^{(2)}(\vec p-\vec q;t)\ 
{\cal C}^{(2)}(\vec p;t_H-t)\ }\ ,
\eea
either by doing it {\sf numerically} (from which we get our central results), or by replacing ${\cal C}^{(2)}(\vec p;t)$ by its {\sf analytic} expression (\ref{disp}). Since the decay (\ref{def2}) has never been studied 
on the lattice, we illustrate the typical situation in Fig.~\ref{fig1}. The two methods give practically indistinguishable results. 
From the matrix elements, for each combination of heavy and light quarks we extract the form factors. For a fixed heavy flavor ($\kappa_h$), one can interpolate (extrapolate) the daughter meson to the kaon-$M_{K}$ (pion-$M_{\pi}$), by using the ``{\sl lattice plane method}'' as discussed in Ref.~\cite{heavy}, which essentially means that one fits each form factor according to the relation:
 \bea
 F_i(\kappa_h,\kappa_\ell) =  \alpha_i + \beta_i M_{P}(\kappa_\ell,\kappa_\ell)^2 +
 \gamma_i M_{P}(\kappa_\ell,\kappa_\ell)^4\nonumber
 \eea
($i\in \{0,+,T\}$). An interesting proposal concerning the extrapolation to the chiral limit is made in Ref.~\cite{lesk}. \\
\begin{figure}[hb]
\vspace*{-.65cm}
\hspace*{-.25cm}
\epsfig{file=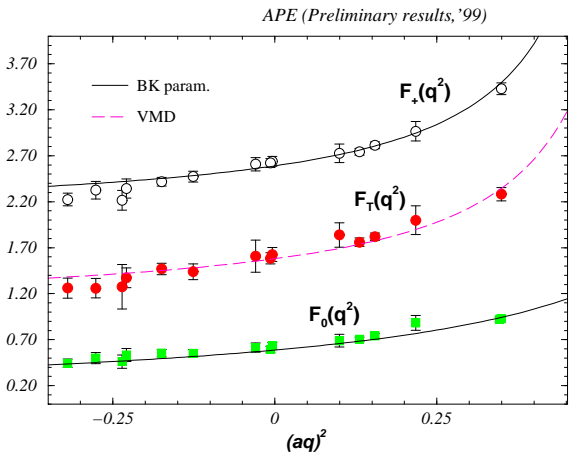,width=7.5cm} 
\vspace*{-1.cm}
\caption{\footnotesize $H\to K\ell^+ \ell^-$ form factors ($F_{+,0}$ also determine $H\to K\ell \nu_\ell$): $F_T$ and $F_+$ are shifted as $1. + F_T(q^2)$ and $2. + F_+(q^2)$.}  
\label{fig3}  
\vspace*{-.8cm}
\end{figure}
\noindent 
The resulting situation is depicted in Fig.~\ref{fig3} where the form factors for $H\to K \ell^+ \ell^-$ are plotted, for $m_H \simeq 2.3$~GeV. 
We stress that this result is not biased by any extrapolation so far. Now we can study their $q^2$-behavior. The recent proposal of Ref.~\cite{BK} is to model form factors $F_+(q^2)$ and $F_0(q^2)$ in such a way that all presently available scaling laws are satisfied. This parametrization (BK) is proposed to avoid the physically unclear notion of double-pole, and to measure the discrepancy with the nearest pole dominance (which is expected to be quite large for the $B$ meson sector, where the kinematically accessible $q^2$-region is large). The parametrization has the form (${\widetilde q}^2=q^2/M_{H^*}^2$)
\bea
{F_0({\widetilde q}^2)} &=& { c_B (1 - \alpha )\over  \   1 - {\textstyle{{\widetilde q}^2}}/{ \textstyle{\large \beta}}  \; } ,\cr
&& \cr
F_+({\widetilde q}^2) &=& { c_B (1 - \alpha )  \over\ \left( 1 - {\widetilde q}^2 \right) \ 
\left( 1 - \alpha {\widetilde q}^2 \right) \; } .
\label{PARAMETR} 
\eea
The physical meaning of parameters is discussed in Ref.~\cite{BK}.
We note that the nearest vector meson dominance\footnote{The vector mesons $M_{H^*}$, which dominate $F_{+,T}(q^2)$, are also computed on the lattice.} for $F_{+,T}(q^2)$, works to an excellent accuracy in the semileptonic region $q^2\in [0, q_{\rm max}^2]$. By including the data at $q^2 < 0$ in the BK-parametrization (as in Fig.2), some small departure from the pole dominance becomes visible~\cite{prepa}. Since our data are straddling around $q^2=0$, which is the point traditionally chosen to quote results, we give the following values (free of any extrapolation in the case of the final $K$ meson):
\bea
&&\hspace*{-6.8mm}F_{0, +}^{D\to K}( 0) = 0.71(3)^{+00}_{-07} ;\ F_{0, +}^{D\to \pi}( 0) =
0.64(5)^{+00}_{-07} ;\cr
&&\cr
&&\hspace*{-6.8mm}F_{T}^{D\to K}( 0) = 0.66(5)^{+00}_{-07} ;\ F_{T}^{D\to \pi }( 0 ) =
0.60(7)^{+00}_{-06} \ , \nonumber \cr
&&\cr
&&\hspace*{-6.8mm}{\rm SU(3)  breaking :}\ F_{0, +}^{D\to K}( 0)/F_{0, +}^{D\to \pi}( 0) = 1.10(4)
 \ , \nonumber
\eea
where the (conservative) systematic uncertainties are determined similarly to those for decay constants. The extrapolation to the $B$-meson introduces a large systematic uncertainty. Namely, at the zero-recoil point, $\vec q = (0,0,0)$ (where the only available form factor is $F_0$), one can use the heavy quark scaling laws. But, once we introduce the recoil (which on our lattice is $|\vec q  |_{\rm min}\sim 700$~MeV), the situation becomes less clear~\cite{prepa}. At $q^2 \to 0$ however, there is a scaling law which stems from the combined heavy quark limit of the decaying, and the large energy limit of the recoiling light meson ($E_P = m_H (1 -
q^2/m_H^2 + m_P^2/m_H^2 )/2$). As discussed in Refs.~\cite{orsay}, all form factors behave as $F_i(0)\simeq m_H^{-3/2}$. After interpolating to $q^2=0$, we fit them as follows~\footnote{Corresponding plots can be obtained upon request from the authors.}:
\bea
\label{scaling}
F_i^{H\to \pi(K)} M_H^{3/2} = {A_i^{\pi(K)} \ \bigl( 1 +
{B_i^{\pi(K)}/ M_H} \bigr) }\ .
\eea
From the fit to this form we extract:
\bea
&&\hspace*{-6.8mm}F_{0 ,+}^{B\to K}( 0) = 0.30(4) ;\ F_{0, +}^{B\to \pi}( 0) =
0.28(4) ; \cr
&&\cr
&&\hspace*{-6.8mm}F_{T}^{B\to K}( 0 ) =  0.29(6) ;\; \ F_{T}^{B\to \pi }( 0 ) = 0.28(7) \ . \nonumber
\eea
Higher ${\cal O}(1/M_H^n)$-corrections to the r.h.s. of (\ref{scaling}) can be large, but they can also be estimated on the lattice, by including more heavy mesons in the simulation. 
The values $F_{0, +}^{B\to \pi}(0)$ agree with those presented at this conference in Ref.~\cite{ukqcd}, which are also obtained by using the NP improvement.


\noindent

\end{document}